\documentclass[11pt]{article}

\usepackage{amsmath}
\usepackage{amssymb}

\usepackage{graphicx}

\usepackage{cite}

\usepackage[labelfont=bf,labelsep=period,justification=raggedright]{caption}
\usepackage{caption}

\makeatletter
\renewcommand{\@biblabel}[1]{\quad#1.}
\makeatother

\date{}

\begin{document}

\begin{flushleft}
{\Large \textbf{Highlighting entanglement of cultures\\
via ranking of multilingual Wikipedia articles} }
\\ \bigskip
Young-Ho\ Eom$^{1}$,
Dima L.\ Shepelyansky$^{1,*}$
\\ \medskip
{1} {\it Laboratoire de Physique Th\'eorique du CNRS, IRSAMC,
Universit\'e de Toulouse, UPS, F-31062 Toulouse, France}
\\
\medskip
$\ast$ E-mail: dima@irsamc.ups-tlse.fr
%$\ast$ Webpage: www.quantware.ups-tlse.fr/dima
\end{flushleft}

\section*{Abstract}

How different cultures evaluate a person? Is an important person
in one culture is also important in the other culture? We address
these questions via ranking of multilingual Wikipedia articles.
With three ranking algorithms based on network structure of
Wikipedia, we assign ranking to all articles in 9 multilingual
editions of Wikipedia and investigate general ranking structure
of PageRank, CheiRank and 2DRank.
In particular, we focus on articles related to persons, identify top
30 persons for each rank among different editions and analyze
distinctions of their distributions over activity fields such as
politics, art, science, religion, sport for each edition. We
find that local heroes are dominant but also global heroes exist
and create an effective network representing
entanglement of cultures.
The Google matrix analysis of network of cultures
shows signs of the Zipf law distribution.
This approach allows to examine diversity and shared
characteristics of knowledge organization between cultures.
The developed computational, data driven approach
highlights   cultural interconnections in a new perspective.\\

\noindent
%{\bf Cite as:  PLoS ONE X(X), eXXX (201X). doi:10.1371/journal.pone.XXX}
Dated: June 26, 2013

\newpage$\phantom{.}$

\section*{Introduction}
%[Cultural diversity]
Wikipedia, the online collaborative encyclopedia, is an amazing
example of human collaboration for knowledge description,
characterization and creation. Like the Library of Babel,
described by Jorge Luis Borges \cite{borges}, Wikipedia goes to
accumulate the whole human knowledge. Since every behavioral
'footprint' (log) is recorded and open to anyone, Wikipedia
provides great opportunity to study various types of social
aspects such as opinion 
consensus~\cite{Kaltenbrunner2012,torok},
language complexity~\cite{Yasseri2012}, and collaboration
structure~\cite{Brandes2009,hecht,nemoto}. 
A remarkable feature of Wikipedia is
its existence in various language editions. In a first
approximation we can attribute each language to an independent
culture, leaving for future refinements of cultures inside one
language. Although Wikipedia has a neutral point of view policy,
cultural bias or reflected cultural diversity is inevitable since
knowledge and knowledge description are also affected by culture
like other human
behaviors~\cite{Norenzayan2011,Gelfand2011,Yasseri2013,Unesco}. 
Thus the cultural
bias of contents~\cite{Callahan2011} becomes an important issue.
Similarity features between various Wikipedia editions
has been discussed at \cite{wang}.
However, the cross-cultural difference between Wikipedia editions
can be also a valuable opportunity for a cross-cultural empirical
study with quantitative approach. Recent steps in this direction,
done for biographical networks of Wikipedia, have been reported in
\cite{aragon2012}.

Here we address the question of  how importance (ranking) of an article in
Wikipedia depends on cultural diversity. In particular, we
consider articles about persons. For instance, is an important
person in English Wikipedia is also important in Korean Wikipedia?
How about French? Since Wikipedia is the product of collective
intelligence, the ranking of articles about persons is a
collective evaluation of the persons by Wikipedia users.
For the ranking of Wikipedia articles we use
PageRank algorithm of Brin and Page \cite{Brin1998},
CheiRank and 2Drank algorithms used in
\cite{Chepelianskii2010,Zhirov2010,2dmotor},
which allow to characterize the information flows
with incoming and outgoing links.
We also analyze the distribution of top ranked persons
over main human activities attributed to
politics, science, art, religion, sport, etc (all others),
extending the approach developed in \cite{Zhirov2010,Eom2013}
to multiple cultures (languages).
The comparison of different cultures shows that they  have
distinct dominance of these activities.

We attribute belongings of top ranked persons at each Wikipedia
language to different cultures (native languages) and in this way
construct the network of cultures. The Google matrix analysis of
this network allows us to find interconnections and entanglement
of cultures. We believe that our computational and statistical
analysis of large-scale Wikipedia networks, combined with
comparative distinctions of different languages, generates novel
insights on cultural diversity.

\section*{Method}
We consider Wikipedia as a network of articles. Each article
corresponds to a node of the network and hyperlinks between
articles correspond to links of the network.
For a given network,
we can define adjacency matrix $A_{ij}$. If there is a link
(one or more quotations) from
node (article) $j$ to node (article)  $i$ then $A_{ij}=1$, 
otherwise, $A_{ij}=0$. The
out-degree $k_{out}(j)$ is the number of links from node $j$ to
other nodes and the in-degree $k_{in}(j)$ is the number of links
to node $j$ from other nodes.

\subsection*{Google matrix}

The matrix $S_{ij}$ of Markov chain transitions
is constructed from adjacency matrix $A_{ij}$
by normalizing sum of elements of each column to unity
($S_{ij}=A_{ij}/\sum_i A_{ij}$,
$\sum_i S_{ij}= 1$) and replacing columns with only zero elements
({\em dangling nodes}) by $1/N$, with $N$ being the matrix size.
Then the  Google matrix of this directed network
has the form \cite{Brin1998,Meyer2006}:
\begin{equation}
   G_{ij} = \alpha  S_{ij} + (1-\alpha)/N \;\; .
\label{eq1}
\end{equation}
In the WWW context the damping parameter $\alpha$
describes the probability
$(1-\alpha)$ to jump to any article (node) for a random walker.
The matrix $G$ belongs to the class of Perron-Frobenius operators, it
naturally appears in dynamical systems \cite{mbrin}.
The right eigenvector at $\lambda = 1$, which is called the PageRank,
has real non-negative elements $P(i)$
and gives a probability $P(i)$ to find a random walker at site $i$.
It is possible to rank all nodes
in a decreasing order of PageRank probability $P(K(i))$
so that the PageRank index $K(i)$ sorts all $N$ nodes
$i$ according their ranks. For large size networks
the PageRank vector and several other eigenvectors
can be numerically obtained using the powerful Arnoldi algorithm
as described in \cite{frahm}. The PageRank vector can be also obtained
by a simple iteration method \cite{Meyer2006}.
Here, we use here the standard value of $\alpha=0.85$ \cite{Meyer2006}.

To rank articles of Wikipedia, we use three ranking algorithms
based on network structure of Wikipedia articles. Detail
description of these algorithms and their use for English Wikipedia
articles are given in ~\cite{Zhirov2010,2dmotor,Eom2013,frahm}.

\subsection*{PageRank algorithm}
PageRank algorithm is originally introduced for Google web search
engine to rank web pages of the World Wide Web (WWW)~\cite{Brin1998}.
Currently PageRank is widely used to rank nodes of network
systems including scientific papers~\cite{Chen2007}, social
network services~\cite{Kwak2010} and even biological
systems~\cite{Kandiah2013}.
Here we briefly outline the iteration method of PageRank computation.
The PageRank vector $P(i,t)$ of a node $i$ at
iteration $ t$ in a network of $N$ nodes is given by

\begin{equation}
P(i,t) = \sum_j G_{ij} P(j,t-1) \; , \;
P(i,t) = (1-\alpha)/N + \alpha\sum_{j} A_{ij}P(j,t-1)/k_{out}(j)  \; .
\label{eq2}
\end{equation}

The stationary state
$P(i)$ of $P(i,t)$ is the PageRank of node $i$. More detail
information about PageRank algorithm is described
in~\cite{Meyer2006}. Ordering all nodes by their decreasing probability
$P(i)$ we obtain the PageRank index $K(i)$.

The essential idea of PageRank algorithm is to use a directed link as
a weighted 'recommendation'. Like in academic citation network, more cited
nodes are considered to be more important. In addition,
recommendations by highly ranked articles
are more important. Therefore high PageRank nodes in the network have
many incoming links from other nodes or incoming links from high
PageRank nodes.

\subsection*{CheiRank algorithm}
While the PageRank algorithm uses information of incoming links to node
$i$, CheiRank algorithm considers information of outgoing links
from node $i$~\cite{Chepelianskii2010,Zhirov2010,2dmotor}. Thus CheiRank is
complementary to PageRank in order to rank nodes in  directed
networks. The CheiRank vector $P^*(i,t)$ of
a node at iteration time $t$ is given by

\begin{equation}
P^*(i) = (1-\alpha)/N + \alpha\sum_{j} A_{ji}P^*(j)/k_{in}(j)
\label{eq3}
\end{equation}
We also  point out that the CheiRank is the right
eigenvector  with maximal eigenvalue $\lambda=1$
satisfying the equation $P^*(i) = \sum_j G^*_{ij} P^*(j)$,
where the Google matrix $G^*$ is built for the network
with inverted directions of links via the standard
definition of $G$ given above.

Like for PageRank, we consider the stationary state $P^*(i)$ of
$P^*(i,t)$ as the CheiRank probability of node $i$ at $\alpha=0.85$. High
CheiRank nodes in the network have a large out-degree.
Ordering all nodes by their decreasing probability
$P^*(i)$ we obtain the CheiRank index $K^*(i)$.

We note that PageRank and CheiRank naturally appear in the world trade network
corresponding to import and export in a commercial exchange between
countries \cite{wtrade}.

The  correlation between PageRank and CheiRank vectors
can be characterized by the correlator
$\kappa$ \cite{Chepelianskii2010,Zhirov2010,2dmotor} defined by
\begin{equation}
\kappa = N \sum_i P(i)P^*(i) -1
\label{eq4}
\end{equation}
The value of correlator for each Wikipedia
edition is represented in Table~\ref{table1}.
All correlators are positive and distributed in the
interval $(1,8)$.

\subsection*{2DRank algorithm}
With PageRank $P(i)$ and CheiRank $P^*(i)$ probabilities,
we can assign PageRank ranking $K(i)$ and CheiRank ranking $K^*(i)$
to each article, respectively. From these two ranks, we can construct
2-dimensional plane of $K$ and $K^*$. The two dimensional ranking
$K_2$ is defined by counting nodes in order of their appearance on
ribs of squares in $(K, K^*)$ plane with the square size growing from
$K=1$ to $K=N$~\cite{Zhirov2010}. A direct detailed
illustration and description of this algorithm is
given in \cite{Zhirov2010}. Briefly, nodes with high
PageRank and CheiRank both get high 2DRank ranking.

\section*{Data description}
We consider 9 editions of Wikipedia including English (EN), French
(FR), German (DE), Italian (IT), Spanish (ES), Dutch (NL), Russian
(RU), Hungarian (HU) and Korean (KO). Since Wikipedia has various
language editions and language is a most fundamental part of culture,
the cross-edition study of Wikipedia
can give us insight on cultural diversity. The
overview summary of parameters of each Wikipedia
is represented in Table
~\ref{table1}.

The corresponding networks of these 9 editions 
are collected and kindly provided to us
by S.Vigna from LAW, Univ. of Milano. 
The first 7 editions in the above list
represent mostly spoken European languages
(except Polish). Hungarian and Korean 
are additional editions representing languages
of  not very large population on European and Asian scales
respectively. They allow us to see interactions not only 
between large cultures but also to see 
links on a small scale. The KO and RU editions allow us to
compare views from European and Asian continents.
We also note that in part these 9 editions reflect the languages
present in the EC NADINE collaboration.

We understand that the present selection of 
Wikipedia editions does represent a complete view
of all 250 languages present at Wikipedia.
However, we think that this selection allows us 
to  perform the  quantitative statistical analysis
of interactions between cultures
making a first step in this direction.

To analyze these interactions we select the fist top 30 
persons (or articles about persons)
appearing in the top ranking list of each of 9 editions
for 3 ranking algorithms of PageRank, CheiRank and 2DRank.
We select these 30 persons manually analyzing 
each list. We attribute each of 30 persons to one of 
6 fields of human activity: politics, science, art, religion, sport,
and etc (here ``etc'' includes all other activities). 
In addition we attribute each person to one of 9 
selected languages or cultures.
We place persons belonging to other languages
inside the additional culture WR (world)
(e.g. Plato).
Usually a belonging of a person 
to  activity field and language 
is taken from the English Wikipedia article about  
this person. If there is no such English Wikipedia article
then we use an article of a Wikipedia edition
language which is native for such a person.
Usually there is no ambiguity in the distribution over
activities and languages. Thus 
Christopher Columbus is attributed to IT culture
and activity field etc,
since English Wikipedia describes him as 
``italian explorer, navigator, and colonizer''.
By our definition politics includes 
politicians (e.g. Barak Obama), emperors
(e.g. Julius Caesar), kings (e.g. Charlemagne).
Arts includes writers (e.g. William Shakespeare),
singers (e.g. Frank Sinatra), painters (Leonardo da Vinci), architects,
artists, film makers (e.g. Steven Spielberg).
Science includes physicists, philosophers (e.g. Plato),
biologists, mathematicians and others.
Religion includes such persons as Jesus, Pope John Paul II.
Sport includes sportsmen (e.g. Roger Federer).
All other activities are placed in activity etc
(e.g. Christopher Columbus, Yuri Gagarin).
Each person belongs only to one language and one
activity field. There are only a few cases which 
can be questioned, e.g.
Charles V, Holy Roman Emperor who is attributed to ES language
since from early long times he was the king of Spain.
All listings of person distributions over the above categories
are presented at the web page given 
at Supporting Information (SI)
and in 27 tables given in SI.

Unfortunately, we were obliged to construct these distributions
manually following each person individually at the Wikipedia
ranking listings. Due to that we restricted our analysis
only to top 30 persons. We think that this number is sufficiently large
so that the statistical fluctuations do not generate
significant changes. Indeed, we find that  our EN distribution
over field activities is close to the one obtained 
for 100 top persons of English Wikipedia dated by Aug 2009
\cite{Zhirov2010}.

To perform additional tests we use the database of about
250000 person names in English, Italian and Dutch
from the research work \cite{aragon2012}
provided to us by P.Arag\'on and A.Kaltenbrunner.
Using this database we were able to use computerized (automatic)
selection of top 100 persons from the ranking lists
and to compare their distributions over activities and
languages with our case of 30 persons.
The comparison is presented in SI in figures S1,S2,S3.
For these 3 cultures
we find that our top 30 persons data are statistically
stable even if the fluctuations are larger
for CheiRank lists. This is in an agreement 
with the fact that the CheiRank probabilities. related to
the outgoing links, are more fluctuating (see discussion at
\cite{Eom2013}).

Of course, it would be interesting to
extend the computerized analysis of personalities to
a larger number of top persons and larger 
number of languages. However, the database of persons
in various languages  still should be cleaned and checked
and also attribution of persons to various activities 
and languages still requires a significant amount of work.
Due to that we present here our analysis only for 30 top persons.
But we note that by itself it represents an interesting case study
since here we have the most important persons
for each ranking. May be the top 1000 persons would be
statistically more stable but clearly a person at position 30 is 
more important than a one at position 1000.
Thus we think that the top 30 persons already give an interesting 
information on links and interactions between cultures.
This information can be used in future more extended studies
of a larger number of persons and languages.

Finally we note that the language is the primary element of culture
even if, of course, culture is not reduced only to language.
In this analysis we use in a first approximation
an equivalence between  language and culture
leaving for future studies the refinement of this link which is 
of course much more complex. In this approximation we
consider that a person like 
Mahatma Gandhi belongs to EN culture since English
is the official language of India. A more advanced study should 
take into account Hindi Wikipedia edition and
attribute this person to this edition.
Definitely our statistical study is only a first step in Wikipedia
based statistical analysis of network of cultures
and their interactions.  

We note that any person from our top 30 ranking belongs
only to one activity field and one culture. We also define local
heros as those who in a given 
language edition are attributed to this language,
and non-local heros as those who belong in a given edition to
other languages. We use category WR (world) where we place
persons who do not belong to any of our 9 languages
(e.g.  Pope John Paul II belongs to WR since his native
language is Polish). 

\section*{Results}
We investigate ranking structure of articles and identify
global properties of PageRank and CheiRank vectors.
The detailed analysis is done for top 30 persons
obtained from the global list of ranked articles
for each of 9 languages. The
distinctions  and common characteristics
of cultures are analyzed by attributing
top 30 persons in each language to
human activities listed above and to their
native language.

\subsection*{General ranking structure}
\begin{figure*}[!ht]
\begin{center}
\includegraphics[width=0.29\columnwidth,angle=-90]{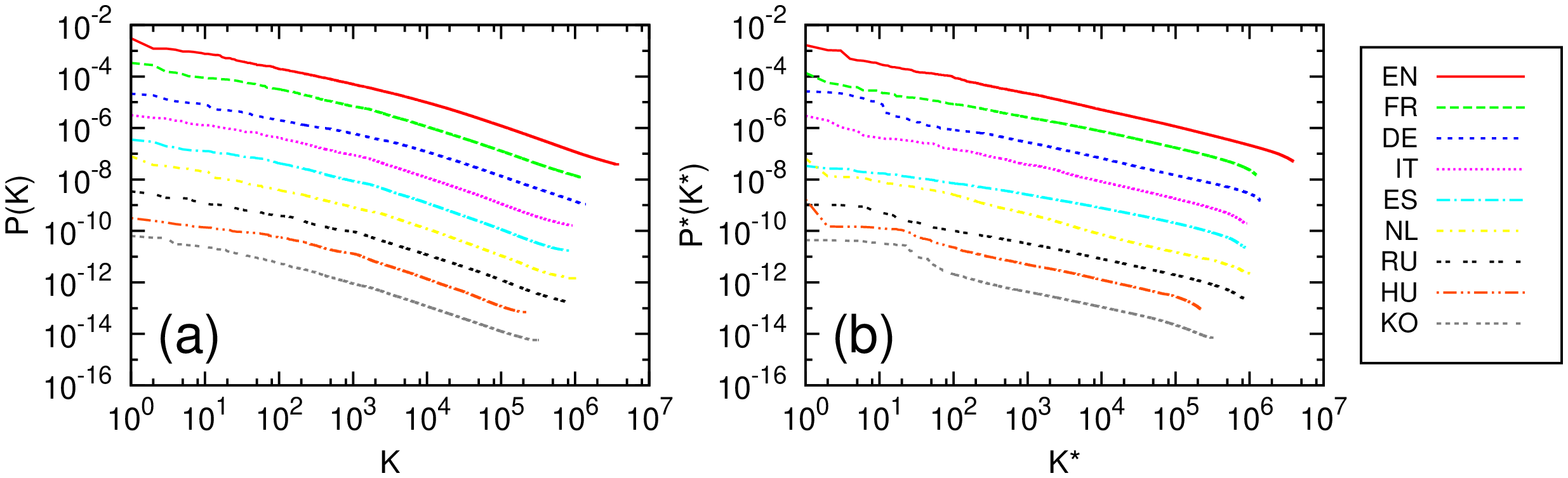}
\caption {\baselineskip 14pt
PageRank probability $P(K)$ as function of
PageRank index $K$ (a) and CheiRank probability
$P^*(K^*)$ as function of CheiRank index
$K^*$ (b). For a better
visualization  each PageRank $P$ and CheiRank $P^*$ curve
is shifted down by a factor
$10^0$ (EN), $10^1$ (FR), $10^2$ (DE), $10^3$ (IT), $10^4$
(ES), $10^5$ (NL), $10^6$ (RU), $10^7$ (HU), $10^8$ (KO).
} \label{fig1}\label{figure1}
\end{center}
\end{figure*}

We calculate PageRank and CheiRank probabilities and indexes
for all networks of considered
Wikipedia editions. The PageRank and CheiRank
probabilities as functions of
ranking indexes are shown in Fig.~\ref{fig1}.
The decay is compatible with an approximate
algebraic decrease of a type $P \sim 1/K^\beta$,
$P^* \sim 1/{K^*}^\beta$ with $\beta \sim 1$
for PageRank and   $\beta \sim 0.6$
for CheiRank.
These values are similar to those
found for the English Wikipedia of 2009
\cite{Zhirov2010}.
The difference of $\beta$ values
originates from
asymmetric nature between in-degree and out-degree
distributions, since PageRank is based on incoming edges
while CheiRank is based on outgoing edges. In-degree
distribution of Wikipedia  editions is broader than out-degree 
distribution of the same edition.
Indeed, the CheiRank probability is proportional to
frequency of outgoing links which has a more rapid decay
compared to incoming one (see discussion in \cite{Zhirov2010}).
The PageRank (CheiRank) probability distributions 
are similar for all editions.
However, the fluctuations of $P^*$
are stronger
that is related  to stronger fluctuations of outgoing edges
\cite{Eom2013}.

The top article of PageRank is usually
{\it USA} or the name of country of a given language
(FR, RU, KO). For NL we have at the top {\it beetle, species, France}.
The top articles of CheiRank are various listings.

\begin{figure*}[!ht]
\begin{center}
\includegraphics[width=0.95\columnwidth]{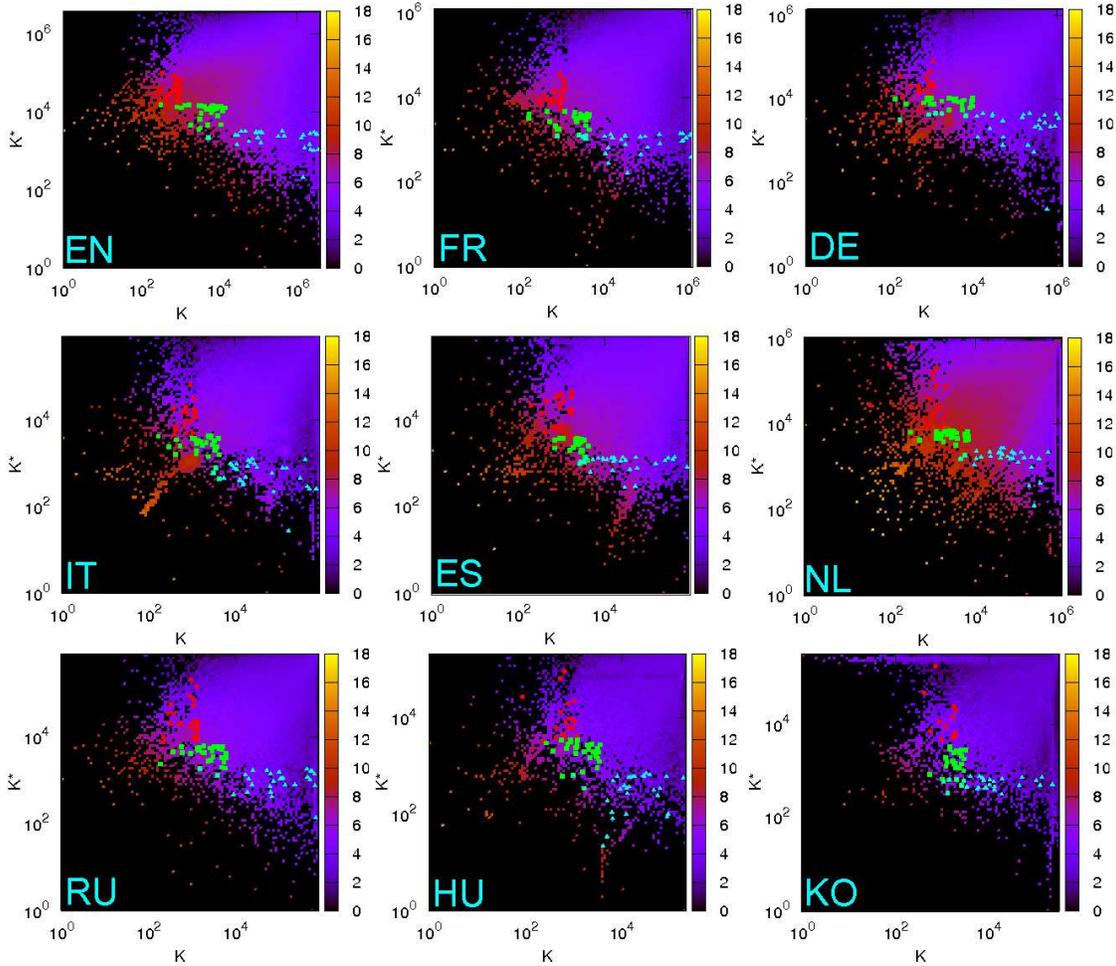}
\caption {\baselineskip 14pt
Density of Wikipedia articles in the
PageRank ranking $K$ versus CheiRank ranking $K^*$ plane for each
Wikipedia edition. The red points are top PageRank articles of
persons, the green points are top 2DRank articles of persons and
the cyan points are top CheiRank articles of persons.
Panels show: English (top-left),
French (top-center), German (top-right),
Italian (middle-left),
Spanish (middle-center), Dutch (middle-left),
Russian (bottom-left),
Hungarian (bottom-center), Korean (bottom-right).
Color bars shown natural logarithm of density, changing
from minimal nonzero density (dark) to maximal one (white),
zero density is shown by black.
} \label{fig2}\label{figure2}
\end{center}
\end{figure*}

Since each article has its PageRank ranking $K$ and CheiRank
ranking $K^*$, we can assign two dimensional coordinates to all
the articles. Fig.~\ref{fig2} shows the density of articles in the two
dimensional plane $(K, K^*)$ for each Wikipedia edition. The
density is computed for $100 \times 100 $ logarithmically
equidistant cells which cover the whole plane $(K, K^*)$. The
density plot represents the locations of articles in the plane. We
can observe high density of articles around
line $K=K^*+const$ that
indicates the positive correlation between PageRank and CheiRank.
However, there are only a few articles within the region of top
both PageRank and CheiRank indexes. We also observe the tendency that
while high PageRank articles ($K<100$) have intermediate CheiRank
($10^2<K^*<10^4)$, high CheiRank articles ($K^*<100$) have broad
PageRank rank values.

\subsection*{Ranking of articles for persons}
We choose top 30 articles about persons for each edition and each
ranking. In Fig.~\ref{fig2}, they are shown by red circles (PageRank),
green squares (2DRank) and cyan triangles (CheiRank). We assign
local ranking $R_{E,A}$ ($1 \ldots 30$) to each person in the list
of top 30 persons for each edition $E$ and ranking algorithm $A$.
An example of $E=EN$ and $A=PageRank$ are given in Table~\ref{table2}.

From the lists of top persons, we identify the "fields" of activity
for each top 30 rank person in which he/she is active on.
We categorize six activity fields - politics,
art, science, religion, sport and etc
(here ``etc'' includes all other activities). As shown in Fig. 3, for
PageRank, politics is dominant and science is secondarily
dominant. The only exception is Dutch where science
is the almost dominant activity field (politics has the same number of points).
In case of 2DRank, art becomes dominant and politics is
secondarily dominant. In case of CheiRank, art and sport are
dominant fields. Thus for example, in CheiRank top 30 list we find
astronomers who discovered a lot of asteroids,
e.g. Karl Wilhelm Reinmuth (4th position in RU
and 7th in DE),
who was a prolific discoverer of about 400 of them.
As a result, his article contains
a long listing of asteroids discovered by him giving him
a high CheiRank.

The change of activity priority for different ranks is due to the
different balance between incoming and outgoing links there.
Usually the politicians are well known for a broad public,
hence, the articles about politicians
are pointed by many articles. However, 
the articles about politician are not very communicative
since they rarely point to other articles.
In contrast, articles about
persons in other fields like science, art  and sport
are more communicative because of
listings of insects, planets, asteroids they discovered, or
listings of song albums or sport competitions they gain.

Next we investigate distributions over "cultures"
to which persons belong. We
determined the culture of person based on
the language the person mainly used
(mainly native language). We consider 10 culture categories - EN, FR, DE, IT,
ES, NL, RU, HU, KO and WR. Here "WR" category represents all other
cultures which do not belong to considered 9 Wikipedia editions.
Comparing with the culture of  persons at various  editions, we can
assign "locality" to each 30 top rank persons for a given Wikipedia
edition and ranking algorithm. For example, as shown in
Table~\ref{table2}, {\it George W. Bush} belongs to "Politics",
"English" and "Local" for English Wikipedia and PageRank, while
{\it Jesus} belongs to ``Religion'', ``World'' WR and ``Non-local''.

As shown in Fig. 4, regardless of ranking algorithms, main part of  top
30 ranking persons of each edition belong to the culture of the
edition (usually about 50\%).
For example, high PageRank persons in English Wikipedia
are mainly English ($53.3 \%$). 
This corresponds to the self-focusing effect discussed in \cite{hecht}.
It is notable that top ranking
persons in Korean Wikipedia are not only mainly Korean ($56.7\%$)
but also the most top ranking non Korean persons in Korean
Wikipedia are Chinese and Japanese ($20\%$). Although there is a
strong tendency that each edition favors its own persons, there is
also overlap between editions. For PageRank, on average, $23.7$
percent of top persons are overlapping while for CheiRank , the
overlap is quite low, only $1.3$ percent. For 2DRank, the overlap is
$6.3$ percent. The overlap of list of top persons implies the
existence of cross-cultural 'heroes'.

To understand the difference between local and non-local top persons
for each edition quantitatively, we consider the PageRank case
because it has a large fraction of non-local top persons. From
Eq.~(\ref{eq2}), a citing article $j$ contributes $\langle
P(j)/k_{out}(j) \rangle$ to PageRank of a node $i$. So the
PageRank $P(i)$ can be high if the node $i$ has many incoming
links from citing articles $j$ or it has incoming links from high
PageRank nodes $j$ with low out-degree $k_{out}(j)$. Thus we can
identify origin of each top person's PageRank using the average
PageRank contribution $\langle P(j)/k_{out}(j) \rangle$ by nodes
$j$ to person $i$ and average number of incoming edges (in-degree)
$k_{in}(i)$ of person $i$ .

As represented in Table~\ref{table3}, considering median, local
top persons have more incoming links than non-local top persons
but the PageRank contribution of the corresponding links are lower
than links of non-local top persons. This indicates that local top
persons are cited more than non-local top persons but non-local
top persons are cited more high weighted links (i.e. cited by
important articles or by articles which don't have many citing
links).

\subsection*{Global and local heroes}
Based on cultural dependency on rankings of persons, we can
identify global and local heroes in the considered Wikipedia
editions. However, for CheiRank  the overlap is very low
and our statistics is not sufficient for selection of
global heroes. Hence we consider only PageRank and 2DRank
cases. We determine the local heroes for each ranking and for
each edition as top persons of the given ranking who belongs to the
same culture as  the edition. Top 3 local heroes for each ranking
and each edition are represented in Table~\ref{table4} (PageRank),
Table~\ref{table5} (CheiRank) and Table~\ref{table6} (2DRank),
respectively.

In order to identify the global heroes, we define ranking score $
\Theta_{P,A} $ for each person $P$ and each ranking algorithm $A$.
Since every person in the top person list has relative ranking
$R_{P,E,A}$ for each Wikipedia edition $E$ and ranking algorithm
$A$ (For instance, in Table~\ref{table2},
$R_{Napoleon,EN,PageRank}=1$). The ranking score $\Theta_{P,A}$
of a person $P$ is give by

\begin{equation}
\Theta_{P,A} = \sum_{E} (31-R_{P,E,A})
\label{eq5}
\end{equation}

According to this definition, a person who appears more often in
the lists of editions and has top ranking in the list gets high
ranking score. We sort this ranking score for each  algorithm.
In this way obtain a list of global heroes for each algorithm. The result is
shown in Table~\ref{table7}. Napoleon is the 1st global hero by
PageRank and Micheal Jackson is the 1st global hero by 2DRank.

\subsection*{Network of cultures}

To characterize the entanglement and interlinking of cultures
we use the data of Fig.~\ref{fig4} and from them construct
the network of cultures.
The image of networks obtained from
top 30 persons of PageRank and 2DRank listings are
shown in Fig.~\ref{fig5} (we do not consider CheiRank case due to
small overlap of persons resulting in a small data statistics).
The weight of directed
Markov transition, or number of links,
from a culture $A$ to a culture $B$ is given by a number of
persons of a given culture $B$ (e.g FR)
appearing in the list of top 30 persons of PageRank (or 2DRank)
in a given culture $A$ (e.g. EN). Thus e.g. for transition
from EN to FR in PageRank we find   $2$ links
(2 French persons in PageRank top 30 persons of English Wikipedia);
for transition from FR to EN in PageRank we have
 $3$ links (3 English persons in PageRank top 30 persons
of French Wikipedia).
The transitions inside
each culture (persons of the same  language as language edition)
are omitted since we are analyzing the interlinks between cultures.
Then the Google matrix of cultures is constructed by the standard rule
for the directed networks:
all links are treated democratically with the same weight,
sum of links in each column is renormalized to unity,
$\alpha=0.85$.  Even if this network has only 10 nodes we still can find
for it PageRank and CheiRank probabilities $P$ and $P^*$
and corresponding indexes $K$ and $K^*$. The matrix elements
of $G$ matrix, written in order of
index $K$, are shown in Fig.~\ref{fig6}
for the corresponding networks of cultures presented in  Fig.~\ref{fig5}.
We note that we consider all cultures on equal democratic grounds.

The decays of PageRank and CheiRank probabilities with the indexes $K, K^*$
are shown in Fig.~\ref{fig7}
for the culture networks of Fig.~\ref{fig5}.
On a first glance a power decay like 
the Zipf law \cite{zipf} $P \sim 1/K$
looks to be satisfactory.
The formal power law fit $P \sim 1/K^z, P^* \sim 1/{(K^*)}^{z^*}$,
done in $\log-\log$-scale for $1\leq K,K^* \leq 10$, 
gives the exponents
$z=0.85 \pm 0.09 , z^*=0.45 \pm 0.09$ (Fig.~\ref{fig7}a),
$z=0.88 \pm 0.10 , z^*=0.77 \pm 0.16$ (Fig.~\ref{fig7}b).
However, the error bars for these fits
are relatively large. Also other statistical tests
(e.g. the Kolmogorov-Smirnov test, see details in \cite{newmantest}) 
give  low statistical accuracy
(e.g. statistical probability $p \approx 0.2; 0.1$ 
and $p \approx 0.01; 0.01 $ for exponents $z,z^*=0.79, 0.42$
and $0.75, 0.65$ in Fig.~\ref{fig7}a
and Fig.~\ref{fig7}b respectively). It is clear that 10 cultures is too
small to have a good statistical accuracy.
Thus,  a larger number of cultures should be used 
to check the validity of the 
generalized Zipf law with a certain exponent. 
We make a conjecture that the Zipf law 
with the generalized exponents
$z, z^*$ will
work in a better way for a larger number of
multilingual Wikipedia editions which
 now have about 250 languages.

The distributions of cultures on the PageRank - CheiRank plane $(K,K^*)$
are shown in Fig.~\ref{fig8}. For the network of cultures
constructed from top 30 PageRank persons
we obtain the following ranking.
The node WR is located at the
top PageRank $K=1$ and it stays at the last CheiRank position
$K^*=10$. This happens due to the fact that such persons
as {\it Carl Linnaeus, Jesus, Aristotle, Plato, Alexander the Great, Muhammad}
are not native for our 9 Wikipedia editions
so that we have many nodes pointing to WR node, while
WR has no outgoing links. The next node in PageRank is FR node
at $K=2, K^*=5$, then DE node at $K=3, K^*=4$
and only then we find EN node at $K=4, K^*=7$.
The node EN is not at all at top PageRank positions
since it has many American politicians
that does not count for links between cultures.
After the world WR the top position is taken by French (FR)
and then German (DE) cultures
which have strong links inside the continental Europe.

However, the ranking is drastically changed
when we consider top 30 2DRank persons.
Here, the dominant role is played by art and science
with singers, artists and scientists.
The world WR here remains at the same position
at $K=1, K^*=10$ but then we obtain
English  EN ($K=2,K^*=1)$ and German DE ($K=3, K^*=5$)
cultures while FR is moved to $K=K^*=7$.

\section*{Discussion}

We investigated cross-cultural diversity of Wikipedia via ranking
of Wikipedia articles. Even if the used ranking algorithms are
purely based on network structure of Wikipedia articles, we find
cultural distinctions and entanglement of
cultures obtained from the multilingual editions of Wikipedia.

In particular, we analyze raking of articles about persons and
identify activity field of persons and cultures to which persons
belong. Politics is dominant in top PageRank persons, art is
dominant in top 2DRank persons and in top CheiRank persons  art
and sport are dominant. We find that
each Wikipedia edition favors its own persons, who have same
cultural background, but there are also cross-cultural non-local heroes,
and even ``global heroes". We establish that local heroes are cited
more often but non-local heroes on average
are cited by more important articles.

Attributing top persons of the ranking list to different cultures
we construct the network of cultures and characterize entanglement
of cultures on the basis of Google matrix analysis of this
directed network.

We considered only 9 Wikipedia editions
selecting top 30 persons in a ``manual'' style.
It would be useful to analyze a larger number of editions
using an automatic computerized selection of persons
from prefabricated listing in many languages
developing lines discussed in \cite{aragon2012}.
This will allow to analyze a large number of persons
improving the statistical accuracy of links between different cultures.

The importance of understanding of cultural diversity in
globalized world is growing. Our computational, data driven
approach can provide a quantitative and efficient way to
understand diversity of cultures by using data created by millions
of Wikipedia users. We believe that our results shed a new light
on how organized interactions and  links between different cultures.

\section*{Acknowledgments}
We thank Sebastiano Vigna \cite{vigna} who kindly provided to us
the network data of 9 Wikipedia editions, collected
in the frame of FET NADINE project. We thank
Pablo Arag\'on and Andreas Kaltenbrunner
for the list of persons in EN, IT, NL which
we used to obtain supporting Figs.S1,S2,S3.
%This research is supported in part by the EC FET Open project
%``New tools and algorithms for directed network analysis''
%(NADINE $No$ 288956).

\section*{Supporting Information}

Supporting Information file
presents Figures S1, S2, S3 showing comparison between probability
distributions over activity fields and language for top 30 and 100
persons for EN, IT, NK respectively;
tables S1, S2, ... S27
showing top 30 persons in PageRank, CheiRank and 2DRank for
all 9 Wikipedia editions. All names are given in English.
Supplementary methods, tables, ranking lists
and figures are available at
http://www.quantware.ups-tlse.fr/QWLIB/wikiculturenetwork/;
data sets of 9 hyperlink networks
are available at \cite{vigna} by a direct request 
addressed to
S.Vigna.

\newpage$\phantom{.}$

\begin{figure*}[!ht]
\begin{center}
\includegraphics[width=0.9\columnwidth]{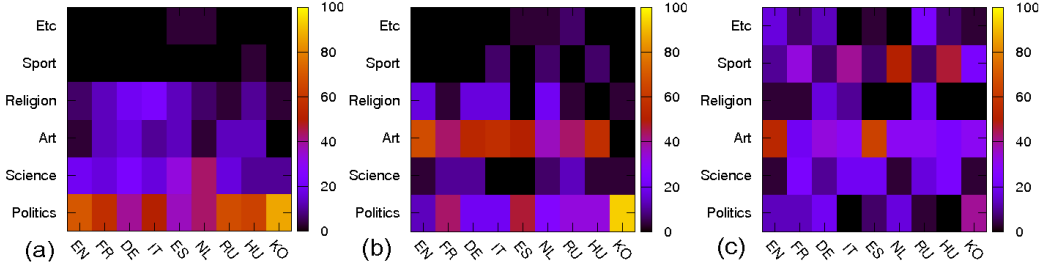}
\caption {\baselineskip 14pt
Distribution of top 30 persons in each rank
over activity fields for each
Wikipedia edition.
Panels correspond to (a) PageRank,
(b) 2DRank, (3) CheiRank.
The color bar shows the values in percents.
} \label{fig3}\label{figure3}
\end{center}
\end{figure*}

\begin{figure*}[!ht]
\begin{center}
\includegraphics[width=0.9\columnwidth]{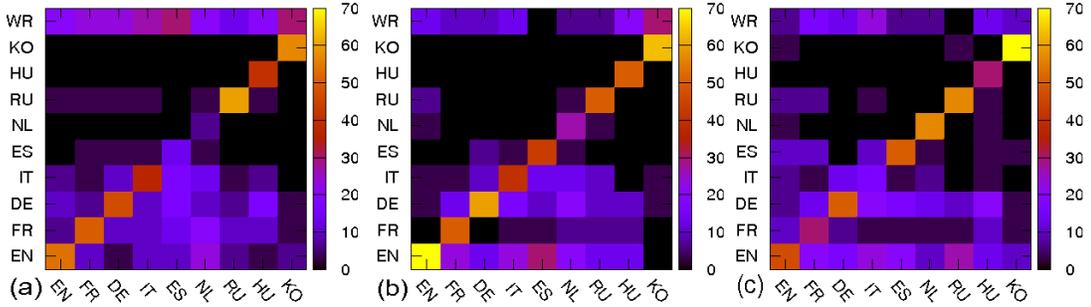}
\caption {\baselineskip 14pt
Distributions of top 30 persons over
different cultures corresponding to
Wikipedia editions,  "WR"
category represents all other cultures
which do not belong to
considered 9 Wikipedia editions.
Panels show ranking by  (a) PageRank, (b) 2DRank, (3)
CheiRank.
The color bar shows the values in percents.
} \label{fig4}\label{figure4}
\end{center}
\end{figure*}

\begin{figure*}[!ht]
\begin{center}
\includegraphics[width=0.9\columnwidth]{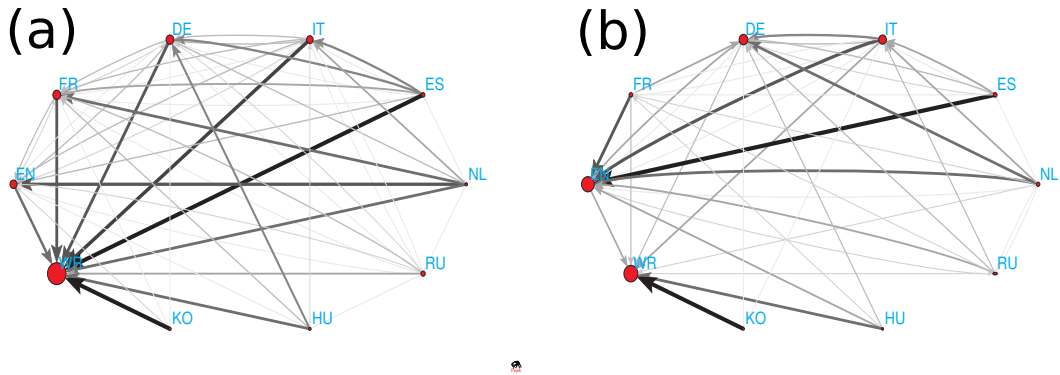}
\caption {\baselineskip 14pt
Network of cultures obtained from 9 Wikipedia languages and
the remaining world (WR) selecting 30 top persons
of PageRank (a) and 2DRank (b) in each culture. The link width
and darkness are proportional to a number of foreign persons quoted
in top 30 of a given culture, the link direction
goes from a given culture to cultures of quoted foreign persons,
quotations inside cultures are not considered.
The size of nodes is proportional to their PageRank.
} \label{fig5}\label{figure5}
\end{center}
\end{figure*}

\begin{figure*}[!ht]
\begin{center}
\includegraphics[width=0.9\columnwidth]{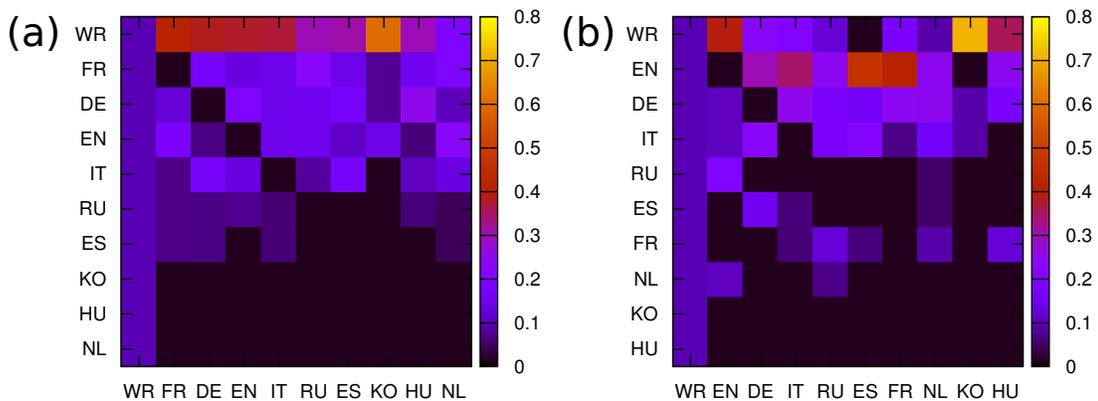}
\caption {\baselineskip 14pt
Google matrix of network of cultures from Fig.~\ref{fig5},
shown respectively for panels $(a), (b)$.
The matrix elements $G_{ij}$ are shown by color
at the damping factor $\alpha=0.85$,
index $j$ is chosen as the PageRank index $K$
of PageRank vector so that the top cultures
with $K=K'=1$ are located at
the top left corner of the matrix.
} \label{fig6}\label{figure6}
\end{center}
\end{figure*}

\begin{figure*}[!ht]
\begin{center}
\includegraphics[width=0.35\columnwidth,angle=-90]{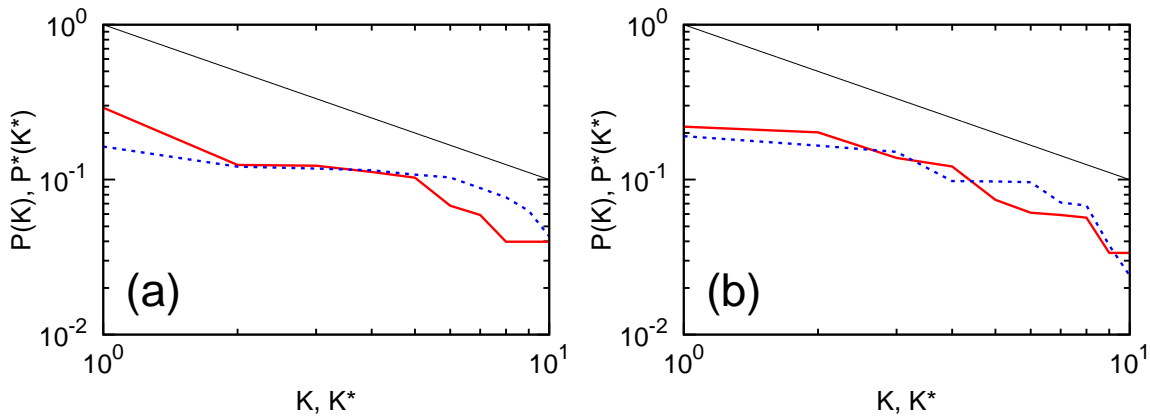}
\caption {\baselineskip 14pt
Dependence of probabilities
of PageRank $P$ (red) and CheiRank $P^*$ (blue) on corresponding
indexes $K$ and $K^*$. The probabilities
are obtained from the network and Google matrix of cultures
shown in Fig.~\ref{fig5} and Fig.~\ref{fig6}
for corresponding panels $(a), (b)$.
The straight lines indicate the Zipf law $P \sim 1/K; P^* \sim 1/K^*$.
} \label{fig7}\label{figure7}
\end{center}
\end{figure*}

\begin{figure*}[!ht]
\begin{center}
\includegraphics[width=0.9\columnwidth]{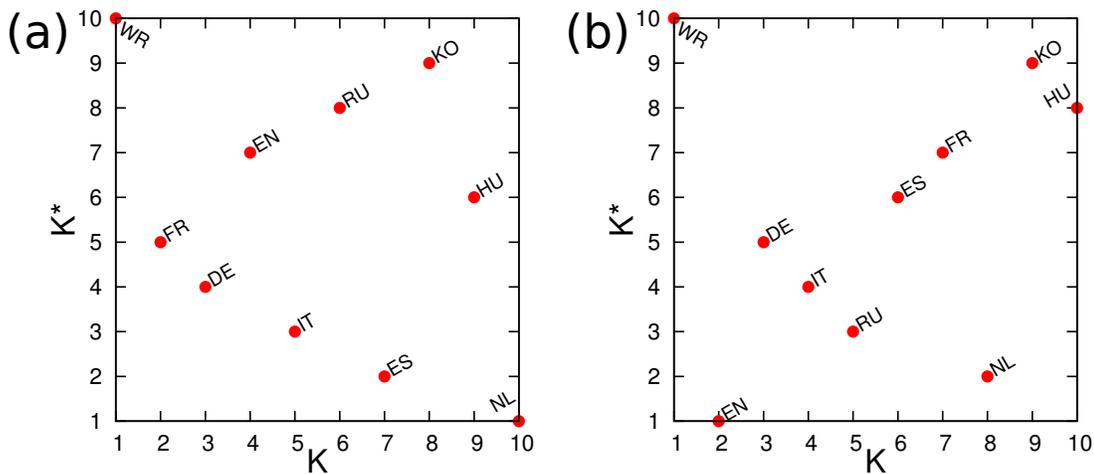}
\caption {\baselineskip 14pt
PageRank versus CheiRank plane of cultures
with corresponding indexes $K$ and $K^*$
obtained from the network of cultures
for corresponding panels $(a), (b)$.
} \label{fig8}\label{figure8}
\end{center}
\end{figure*}

\newpage$\phantom{.}$

\begin{table}[!ht]
\caption{Considered Wikipedia networks from language editions:
English (EN), French (FR), German (DE),
Italian (IT), Spanish (ES), Dutch (NL), Russian (RU),
Hungarian (HU), Korean (KO). Here $N_A$ is number of articles,
$N_L$ is number of hyperlinks between articles,
$\kappa$ is the correlator between PageRank and CheiRank.
Date represents the time in which
data are collected. }
\begin{center}
\resizebox{8cm}{!}{
\begin{tabular}{|c|c|c|c|c|}
  \hline
  Edition & $N_{A}$ & $N_{L}$ & $\kappa$  & Date \\
  \hline
EN & 3920628 & 92878869 & 3.905562 & Mar. 2012 \\
FR & 1224791 & 30717338 & 3.411864 & Feb. 2012 \\
DE & 1396293 & 32932343 & 3.342059 & Mar. 2012 \\
IT & 917626  & 22715046 & 7.953106 & Mar. 2012 \\
ES & 873149  & 20410260 & 3.443931 & Feb. 2012 \\
NL & 1034912 & 14642629 & 7.801457 & Feb. 2012 \\
RU & 830898  & 17737815 & 2.881896 & Feb. 2012 \\
HU & 217520  & 5067189  & 2.638393 & Feb. 2012 \\
KO & 323461  & 4209691  & 1.084982 & Feb. 2012 \\
  \hline
\end{tabular}}
\end{center}
\label{table1}
\end{table}

\begin{table}[!ht]
\caption{Example of list of top 10 persons by PageRank for English
Wikipedia with their field of activity and native language. }
\begin{center}
\resizebox{13cm}{!}{
\begin{tabular}{|c|c|c|c|c|}
  \hline
  $R_{EN,PageRank} $ & Person & Field & Culture & Locality \\
  \hline
1 & Napoleon & Politics & FR & Non-local \\
2 & Carl Linnaeus & Science & WR & Non-local \\
3 & George W. Bush & Politics & EN & Local \\
4 & Barack Obama & Politics & EN & Local \\
5 & Elizabeth II & Politics & EN & Local \\
6 & Jesus & Religion & WR & Non-local \\
7 & William Shakespeare & Art & EN & Local\\
8 & Aristotle & Science & WR & Non-local \\
9 & Adolf Hitler & Politics & DE & Non-local \\
10 & Bill Clinton & Politics & EN & Local \\
  \hline
\end{tabular}}
\end{center}
\label{table2}
\end{table}

\begin{table*}[!ht]
\caption{PageRank contribution per link and in-degree of
PageRank local and non-local heroes $i$ for each edition. 
$ [P(j)/k(j)_{out} ]_{L}$ and $ [ P(j)/k(j)_{out} ]_{NL}$ are median
PageRank contribution of a local hero $L$ and non-local hero $NL$
by a article $j$ which cites local heroes $L$ and non-local heroes
$NL$ respectively. $ [k(L)_{in}]$ and $ [k(NL)_{in}]$ are median
number of in-degree $k(L)_{in}$ and $k(NL)_{in}$ of local
hero $L$ and non-local hero $NL$, respectively. $N_{Local}$ is
number local heroes in given edition.}
\begin{center}
\resizebox{15cm}{!}{
\begin{tabular}{|c|c|c|c|c|c|c|c|}
  \hline
  Edition & $N_{Local}$ & $[P(j)/k(j)_{out}]_{L}$ & & $[P(j)/k(j)_{out}]_{NL}$ & $[k(L)_{in}]$ & &$[k(NL)_{in}]$ \\
  \hline
EN & 16 & $1.43\times 10^{-8}$ & $<$ & $2.18\times 10^{-8}$ & $5.3\times 10^{3}$ & $>$ &$3.1\times 10^{3}$ \\
FR & 15 & $3.88\times 10^{-8}$ & $<$ & $5.69\times 10^{-8}$ & $2.6\times 10^{3}$ & $>$ &$2.0\times 10^{3}$ \\
DE & 14 & $3.48\times 10^{-8}$ & $<$ & $4.29\times 10^{-8}$ & $2.6\times 10^{3}$ & $>$ &$2.1\times 10^{3}$ \\
IT & 11 & $7.00\times 10^{-8}$ & $<$ & $7.21\times 10^{-8}$ & $1.9\times 10^{3}$ & $>$ &$1.5\times 10^{3}$ \\
ES & 4  & $5.44\times 10^{-8}$ & $<$ & $8.58\times 10^{-8}$ & $2.2\times 10^{3}$ & $>$ &$1.2\times 10^{3}$ \\
NL & 2  & $7.77\times 10^{-8}$ & $<$ & $14.4\times 10^{-8}$ & $1.0\times 10^{3}$ & $>$ &$6.7\times 10^{2}$ \\
RU & 18 & $6.67\times 10^{-8}$ & $<$ & $10.2\times 10^{-8}$ & $1.7\times 10^{3}$ & $>$ &$1.5\times 10^{3}$ \\
HU & 12 & $21.1\times 10^{-8}$ & $<$ & $32.3\times 10^{-8}$ & $8.1\times 10^{2}$ & $>$ &$5.3\times 10^{2}$ \\
KO & 17 & $16.6\times 10^{-8}$ & $<$ & $35.5\times 10^{-8}$ & $4.7\times 10^{2}$ & $>$ &$2.3\times 10^{2}$ \\
  \hline
\end{tabular}}
\end{center}
\label{table3}
\end{table*}

\begin{table*}[!ht]
\caption{List of local heroes by PageRank for each Wikipedia
edition. All names are represented by article titles in English
Wikipedia. Here "William the Silent" is the third local hero in Dutch
Wikipedia but he is out of top 30 persons.} \small
\begin{center}
\resizebox{17cm}{!}{
\begin{tabular}{|c|c|c|c|}
  \hline
  Edition & 1st & 2nd & 3rd \\
  \hline
EN & George W. Bush & Barack Obama & Elizabeth II \\
FR & Napoleon & Louis XIV of France & Charles de Gaulle \\
DE & Adolf Hitler & Martin Luther & Immanuel Kant \\
IT & Augustus & Dante Alighieri & Julius Caesar \\
ES & Charles V, Holy Roman Emperor & Philip II of Spain & Francisco Franco \\
NL & William I of the Netherlands & Beatrix of the Netherlands & William the Silent \\
RU & Peter the Great  & Joseph Stalin & Alexander Pushkin \\
HU & Matthias Corvinus  & Szent\'agothai J\'anos  & Stephen I of Hungary \\
KO & Gojong of the Korean Empire & Sejong the Great & Park Chung-hee  \\
  \hline
\end{tabular}}
\end{center}
\label{table4}
\end{table*}

\begin{table*}[!ht]
\caption{List of local heroes by CheiRank for each Wikipedia
edition. All names are represented by article titles in English
Wikipedia. } \scriptsize
\begin{center}
\resizebox{17cm}{!}{
\begin{tabular}{|c|c|c|c|}
  \hline
  Edition & 1st & 2nd & 3rd \\
  \hline
EN & C. H. Vijayashankar & Matt Kelley & William Shakespeare (inventor) \\
FR & Jacques Davy Duperron & Jean Baptiste Ebl\'e & Marie-Magdeleine Aym\'e de La Chevreli\`{e}re \\
DE & Harry Pepl & Marc Zwiebler & Eugen Richter \\
IT & Nduccio & Vincenzo Olivieri & Mina (singer) \\
ES & Che Guevara & Arturo Mercado & Francisco Goya \\
NL & Hans Renders & Julian Jenner & Marten Toonder \\
RU & Aleksander Vladimirovich  Sotnik & Aleksei Aleksandrovich Bobrinsky & Boris Grebenshchikov \\
HU & Csernus Imre  & Kati Kov\'acs  & Pl\'eh Csaba \\
KO & Lee Jong-wook (baseball) & Kim Dae-jung & Kim Kyu-sik \\
  \hline
\end{tabular}}
\end{center}
\label{table5}
\end{table*}

\begin{table*}[!ht]
\caption{List of local heroes by 2DRank for each Wikipedia
edition. All names are represented by article titles in English
Wikipedia}
\begin{center}
\resizebox{16cm}{!}{
\begin{tabular}{|c|c|c|c|}
  \hline
  Edition & 1st & 2nd & 3rd \\
  \hline
EN & Frank Sinatra & Paul McCartney & Michael Jackson \\
FR & Fran\c{c}ois Mitterrand & Jacques Chirac & Honor\'e de Balzac \\
DE & Adolf Hitler & Otto von Bismarck & Ludwig van Beethoven \\
IT & Giusppe Garibaldi & Raphael & Benito Mussolini \\
ES & Sim\'on Bol\'ivar & Francisco Goya & Fidel Castro \\
NL & Albert II of Belgium & Johan Cruyff & Rembrandt \\
RU & Dmitri Mendeleev & Peter the Great  & Yaroslav the Wise \\
HU & Stephen I of Hungary  & S\'andor Pet\H{o}fi  & Franz Liszt \\
KO & Gojong of the Korean Empire &  Sejong the Great & Park Chung-hee  \\
  \hline
\end{tabular}}
\end{center}
\label{table6}
\end{table*}

\begin{table*}[!ht]
\caption{List of global heroes by PageRank and 2DRank for all 9
Wikipedia editions. All names are represented by article titles in
English Wikipedia. Here, $\Theta_A$ is the ranking score of the
algorithm $A$ (\ref{eq5});
$N_A$ is the number of appearances of a given person
in the top 30 rank for all editions.}
\begin{center}
\resizebox{16cm}{!}{
\begin{tabular}{|c|c|c|c|c|c|c|c|c|}
  \hline
 Rank & PageRank global heroes & $\Theta_{PR}$ & $N_A$ & 2DRank global heroes & $\Theta_{2D}$ & $N_A$\\
  \hline
1st & Napoleon & 259 & 9 & Micheal Jackson & 119 & 5 \\
2nd & Jesus & 239 & 9 & Adolf Hitler & 93 & 6 \\
3rd & Carl Linnaeus & 235 & 8 & Julius Caesar & 85 & 5 \\
4th & Aristotle & 228 & 9 & Pope Benedict XVI & 80 & 4 \\
5th & Adolf Hitler & 200 & 9 & Wolfgang Amadeus Mozart & 75 & 5 \\
6th & Julius Caesar & 161 & 8 & Pope John Paul II & 71 & 4 \\
7th & Plato & 119 & 6 & Ludwig van Beethoven & 69 & 4 \\
8th & Charlemagne & 111 & 8 & Bob Dylan & 66 & 4 \\
9th & William Shakespeare & 110 & 7 & William Shakespeare & 57 & 3 \\
10th & Pope John Paul II & 108 & 6 & Alexander the Great & 56 & 3 \\
  \hline
\end{tabular}}
\end{center}
\label{table7}
\end{table*}

\end{document}